\documentclass[english,a4paper]{article}

\usepackage{bm}
\usepackage{graphicx}
\usepackage{amsmath}
\usepackage{amsfonts}
\usepackage{amssymb}
\usepackage{amsthm}
\usepackage{amstext}
\usepackage{amsbsy}
\usepackage{amsopn}
\usepackage{amscd}
\usepackage{amsxtra}
\usepackage[colorlinks=true,allcolors=blue]{hyperref}
\usepackage{color}
\usepackage{soul}
\usepackage[dvipsnames]{xcolor}
\usepackage{listings}

\usepackage{cite}

\begin{document}

\title{Optimal control of a Bose-Einstein Condensate in an optical lattice: The non-linear and two-dimensional cases}

\author{E. Dionis, B. Peaudecerf\footnote{Laboratoire Collisions Agr\'egats R\'eactivit\'e, UMR 5589, FeRMI, UT3, Universit\'e de Toulouse, CNRS, 118 Route de Narbonne, 31062 Toulouse CEDEX 09, France}, S. Gu\'erin\footnote{Laboratoire Interdisciplinaire Carnot de Bourgogne, CNRS UMR 6303, Universit\'{e} de Bourgogne, BP 47870, F-21078 Dijon, France}, D. Gu\'ery-Odelin\footnote{Laboratoire Collisions Agr\'egats R\'eactivit\'e, UMR 5589, FeRMI, UT3, Universit\'e de Toulouse, CNRS, 118 Route de Narbonne, 31062 Toulouse CEDEX 09, France}, D. Sugny\footnote{Laboratoire Interdisciplinaire Carnot de Bourgogne, CNRS UMR 6303, Universit\'{e} de Bourgogne, BP 47870, F-21078 Dijon, France, dominique.sugny@u-bourgogne.fr}}

\maketitle
\begin{abstract}
We numerically study the optimal control of an atomic Bose-Einstein condensate in an optical lattice. We present two generalizations of the gradient-based algorithm, GRAPE, in the non-linear case and for a two-dimensional lattice. We show how to construct such algorithms from Pontryagin's maximum principle. A wide variety of target states can be achieved with high precision by varying only the laser phases setting the lattice position. We discuss the physical relevance of the different results and the future directions of this work.
\end{abstract}

\section{Introduction}
Quantum technologies seek to exploit the specific properties of quantum systems for real-world applications in computing, sensing, simulations or communications~\cite{acin18}. In this framework, quantum optimal control (QOC) can be viewed as a set of methods for designing and implementing external electromagnetic fields to realize specific operations on a quantum device in the best possible way~\cite{glaserreview,brifreview,roadmap}. QOC is becoming a key tool in many different experimental platforms, ranging from superconducting circuits~\cite{wilhelm2021,abdel2020,review} to cold atoms~\cite{borzi2007,tutorial24,frank16}, molecular physics~\cite{RMPsugny} or NV centers~\cite{NVcenters}. Despite the maturity and effectiveness of optimal control techniques which rely on a rigorous mathematical framework, namely the Pontryagin Maximum Principle~(PMP, see~\cite{tutorial24,pont,boscain21,liberzon} for details), developments and adaptations of standard methods are necessary to take into account experimental limitations or additional degrees of freedom in the experiment~\cite{bryson,grape,reich,gross,AD3,meri2023,dionis2023}. Bose-Einstein condensates (BEC) constitute a promising system~~\cite{eckardt17,bloch08} for applications in quantum sensing and simulation~\cite{gross17} in which optimal control may play a major role to prepare specific states or to improve the estimation of unknown parameters~\cite{roadmap}. This approach has been applied in a variety of works both theoretically and experimentally, with very good agreement~\cite{borzi2007,tutorial24,frank16,dupont21,jager2013,jager2014,rodzinka24,saywell20,sorensen18,hocker16,mennemann15,chen11,zhang16,bucker13,frank14,potting01,weidner17,weidner18,bason12,zhou18,arrouas23,dupont23,amri19,adriazola22,OCTBEC}. A specific example is given by a BEC trapped in an optical lattice. In this case, the two control parameters are usually the depth and  phase of the lattice that can be precisely adjusted experimentally~\cite{dupont21}. However, a majority of studies consider a simplified situation in which the non-linear term of the Gross-Pitaevski equation describing the dynamics of the BEC is neglected. This approximation that can be justified by the low density of the BEC is crucial and allows to study the system in a unitary framework given by the Schr\"odinger equation. It also greatly simplifies the implementation of QOC as recently shown by our group in~\cite{dupont21,dupont23} where a gradient-based algorithm, GRAPE~\cite{grape}, has been used. Different formulations of optimal control algorithms have been proposed in the non-linear case. The optimal control of a BEC in a magnetic microtrap has been investigated numerically in~\cite{borzi2007,jager2013,jager2014}. The control problem has been solved analytically for a two-level quantum system in~\cite{zhang2011,chen2016,dorier2017,zhu2024,zhu2020}. The control of the Gross-Pitaevski  equation has also been the subject of a series of mathematical papers (see~\cite{feng2016,hintermuller2013} to mention a few). In this article, we propose to revisit such works by applying the PMP in the presence of mean-field interactions and deriving the corresponding algorithm from this optimization principle. Intensive use of the pseudo-spectral approach~\cite{littlejohn2002,light1992,leforestier91,kosloff83,guerin99} and FBR-DVR bases (Finite Basis Representation and Discrete Variable Representation) is necessary to analytically express the different quantities and accelerate the numerical calculation. We demonstrate the effectiveness of this algorithm and discuss the role of nonlinearity on the control procedure.

The use of one-dimensional lattices can limit the range of phenomena accessible to quantum simulation: higher dimensionalities e.g. can radically change the physics of localization~\cite{morsch2006}, and also increase the role of interaction, giving access to many-body phenomena such as the Mott transition~\cite{bloch08}.  It is thus of utmost importance to extend the 1D optimal approach to the 2D or 3D cases. This problem has been investigated by very few studies such as~\cite{mennemann15,zhou18}. In~\cite{mennemann15}, optimal control is applied to a BEC in a three dimensional magnetic trap. An optimization algorithm different from GRAPE has been used to control BEC in 2D and 3D lattices in~\cite{zhou18}. We present in this work another numerical implementation of QOC to a 2D lattice. We consider here a triangular geometry, which is non-separable. This example can be used as a test-bed for other geometries and for 3D lattices. State-to-state transfer can be optimized numerically with a very good precision. We discuss on this example the controllability of the target state with respect to the number of independent controls available.

The paper is organized as follows. Section~\ref{sec2} briefly recalls the specifics of the experimental setup and the application of QOC in a 1D and linear cases. Section~\ref{sec3} is dedicated to the extension of the GRAPE algorithm to the non-linear case. Optimal control with two-dimensional lattices is described in Sec.~\ref{sec4}. Conclusion and prospective views are given in Sec.~\ref{sec5}.

\section{The standard one-dimensional case}\label{sec2}
Cold atom systems are characterized by their large size and the broad range of controllable parameters they offer, which makes them excellent candidates for applications in quantum technologies. We consider in this paper the BEC experiment in Toulouse in the group of D. Gu\'ery-Odelin. Recent experimental results have shown the key role of QOC for state-to-state transfer in this setup~\cite{tutorial24,dupont21,arrouas23,dupont23}.

The experiment starts by laser cooling followed by evaporation of a Rubidium 87 gas allowing the formation of a BEC. The condensate is composed of $5\times 10^5$ atoms at a temperature of 90~nK and is held in a hybrid magneto-optical trap to confine the BEC and compensate for gravity. A horizontal one-dimensional optical lattice trap, formed by the interference of two counter propagating laser beams with a wavelength $\lambda=1064$~nm, is superimposed on the hybrid trap, aligned with the dipole trapping beam. Using acousto-optic modulators, the amplitude and phase of the lattice lasers can be shaped in time. This amounts to modifying the depth of the lattice or translating it, respectively. Following the experimental results of~\cite{dupont21,arrouas23,dupont23}, we assume here that the lattice depth is fixed and only the relative phase of the lasers can be controlled.

The wave function $|\psi(t)\rangle$ describing the state of the BEC is governed by the time-dependent Schr\"odinger equation which can be expressed as
\begin{align}
  \imath \hbar \frac{d|\psi(t)\rangle}{dt} = \left( \frac{\hat{p}^2}{2m} - \frac{s E_L}{2}\cos\left( k_L \hat{x} + \varphi(t) \right) \right) |\psi(t)\rangle,
\end{align}
where $\hat{p}$ and $\hat{x}$ are respectively the momentum and the position operators, with $x$ the spatial coordinate along the axis of the optical lattice. We denote by $m$ the atom mass, $k_L=2\pi/(\lambda/2)$ the wave vector and $E_L=\left(\hbar^2 k_L^2\right)/\left(2m\right)$ the characteristic energy of the lattice. The parameters $s$ and $\varphi$ correspond respectively to the relative amplitude of the lattice and to its phase. In standard experimental conditions, we stress that the atom interactions and the potential energy due to the hybrid trap, $V_{\text{hyb}}=\frac{1}{2}m\omega_{\textrm{ext}}x^2$ with $\omega_{\text{ext}}=2\pi \times f$ where $f$ varies from a few Hz to 25~Hz at most, can be neglected due to the low density of the condensate and the short timescale of the atomic dynamics ( $t_\textrm{dyn}\ll 2\pi/\omega_\textrm{ext}$).

In order to work with dimensionless coordinates, we introduce the following change of variables $t \rightarrow \frac{E_L}{\hbar}t$ and $x \rightarrow k_L x$. We obtain
\begin{equation}
  \label{sysBEC1}
  \imath \frac{d|\psi(t)\rangle}{dt} =\left(\hat{p}^2 - \frac{s}{2}\cos\left( \hat{x} + \varphi(t) \right)\right)|\psi(t)\rangle,
\end{equation}
where $\hat{p}=-\imath\frac{\partial}{\partial x}$ and $\hat{x}=x$  in the position representation. Note that the study of this dimensionless system makes it possible to transfer the results of this paper to other experimental setups with different characteristics.
The eigenvectors of the momentum operator are denoted by $|\phi_{\alpha}\rangle$, of wave function in the $x$ representation  $\phi_{\alpha}(x)=\frac{1}{\sqrt{2\pi}}e^{\imath \alpha x}$ and of eigenvalue $\alpha$. Since the potential is periodic in $x$, the Bloch theorem states that  $\alpha = n + q$, where $n \in \mathbb{Z}$ is a relative integer and $q\in [-0.5,0.5]$ is the quasi-momentum. The periodicity of the potential also implies that the quasi-momentum is a constant of the motion during the control process. In a sub-Hilbert space characterized by a specific value of $q$, the state of the system can be expanded in the plane wave basis as follows
       \begin{align}
  |\psi\rangle = \sum_{n\in\mathbb{Z}}c_{q,n} |\phi_{q+n}\rangle.
\end{align}
It is then straightforward to show that the coefficients $c_{q,n}$ are solutions of the following equation
\begin{align}
   \label{BEC_coef} \imath\dot{c}_{q,n} &= \left( n+q \right)^2c_{q,n} - \frac{s}{4}\left(e^{\imath\varphi(t)}c_{q,n-1} + e^{-\imath\varphi(t)}c_{q,n+1} \right).
\end{align}
We deduce that the Schr\"odinger equation can be written in matrix form as
  \begin{align}
   \label{sysBEC}\imath\frac{d|\psi(t)\rangle}{dt} = \hat{H}|\psi(t)\rangle = \left( \hat{H}_0 + \cos\left( \varphi(t)\right)\hat{H}_1 + \sin\left( \varphi(t)\right)\hat{H}_2\right) |\psi(t)\rangle,
\end{align}
with\begin{align}
   |\psi(t)\rangle = \begin{pmatrix}
  \vdots \\
  c_{q,n-1} \\
  c_{q,n} \\
  c_{q,n+1} \\
  \vdots
  \end{pmatrix},
\end{align}
       \begin{align}
       \centering
  \label{BEC_vector} \hat{H}_0 =  \begin{pmatrix}
  & \ddots &  &  &  \\
  \dots & 0 & \left((n-1)+q\right)^2 & 0 & 0 & 0 & \dots \\
  \dots & 0 & 0 & \left(n+q\right)^2 & 0 & 0 & \dots \\
  \dots & 0 & 0 & 0 & \left((n+1)+q\right)^2 & 0 & \dots \\
   &  &  &  &  & \ddots &
  \end{pmatrix},
\end{align}
and
       \begin{align}
       \centering
  \hat{H}_1 = \begin{pmatrix}
  \ddots &  & \ddots &  &  \\
  \dots & -\frac{s}{4} & 0 & -\frac{s}{4} & 0 & 0 & \dots \\
  \dots & 0 & -\frac{s}{4} & 0 & -\frac{s}{4} & 0 & \dots \\
  \dots & 0 & 0 & -\frac{s}{4} & 0 & -\frac{s}{4} & \dots \\
   &  &  &  & \ddots &  & \ddots
  \end{pmatrix}, \
  \hat{H}_2 = \begin{pmatrix}
  \ddots &  & \ddots &  &  \\
  \dots & -\imath\frac{s}{4} & 0 & \imath\frac{s}{4} & 0 & 0 & \dots \\
  \dots & 0 & -\imath\frac{s}{4} & 0 & \imath\frac{s}{4} & 0 & \dots \\
  \dots & 0 & 0 & -\imath\frac{s}{4} & 0 & \imath\frac{s}{4} & \dots \\
   &  &  &  & \ddots &  & \ddots
  \end{pmatrix}.
  \end{align}
From a numerical point of view, the infinite-dimensional Hilbert space is truncated to a finite one such that $|n|\leq n_{\textrm{max}}$, where $n_{\textrm{max}}$ is chosen as a function of the initial and target states in order to avoid edge effects in the computation. In the numerical simulations, we consider generally $n_{\textrm{max}}=10$, which leads to a truncated space of dimension $2\times n_{\textrm{max}}+1=21$. The parameter $n_{\textrm{max}}$ is adjusted in each case depending on the target state.

Different theoretical and experimental implementations of QOC have shown the effectiveness of the procedure and the ability to achieve a variety of target states $|\psi_t\rangle$, with a very good match between theory and experiment~\cite{dupont21,dupont23}. To this aim, the optimal phase is calculated from the application of GRAPE~\cite{tutorial24,grape} to the dynamical system~(\ref{sysBEC}). The goal of the optimal control problem is to maximize the fidelity $F=|\langle\psi_t|\psi(t_f)\rangle|^2$ at time $t_f$. The PMP leads to the Pontryagin Hamiltonian $H_p = \Im(\langle\chi(t)|\hat{H}|\psi(t)\rangle)$, with $|\chi(t)\rangle$ the adjoint state, whose dynamics are also governed by Eq.~(\ref{sysBEC}), with the final condition $\chi_0\langle\psi_t|\psi(t_f)\rangle |\psi_t\rangle$, where $\chi_0$ is the dual variable of the cost. $\chi_0$ is set to 1/2 in the numerical simulation. We refer the interested reader to the recent tutorial~\cite{tutorial24} for a comprehensive introduction to the PMP for quantum optimal control.

\begin{figure}
        \centering
        \includegraphics[width=1\textwidth]{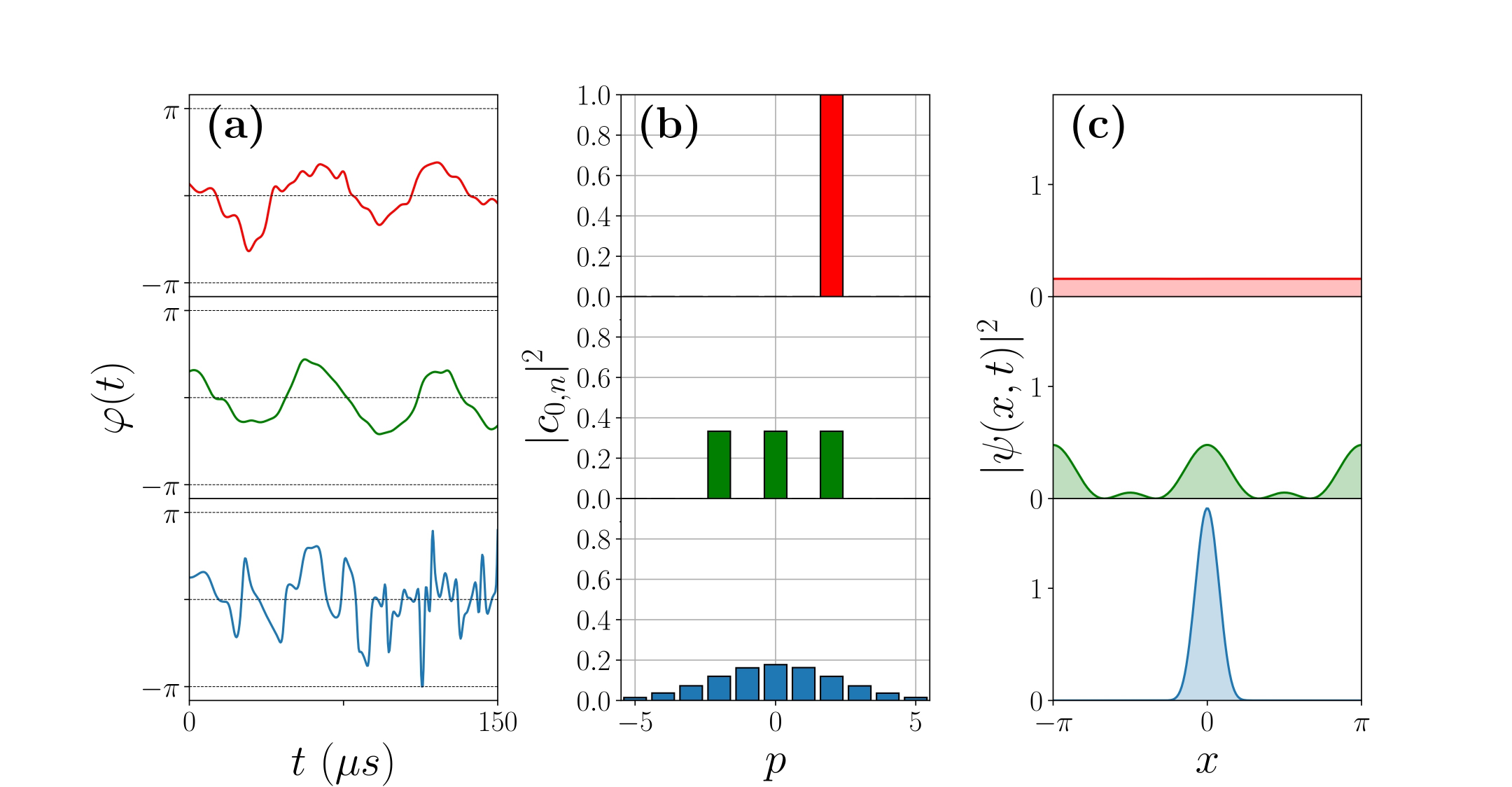}
        \caption{Example of three state-to-state transfers of a BEC from the initial state $|\phi_{0+0}\rangle$. The first, second and third rows (from top to bottom) correspond respectively to the target states $|\phi_{0+2}\rangle$, $\frac{1}{\sqrt{3}}\left(|\phi_{0-2}\rangle +|\phi_{0+0}\rangle + |\phi_{0+2}\rangle\right)$ and $|g(0,0,1/3)\rangle$. The columns (a), (b) and (c) represent respectively the corresponding controls $\varphi(t)$, the final population distribution in the momentum basis and the probability distribution in position at the final time.}
   \label{oct_BEC}
\end{figure}

Using the maximization condition of the PMP, the control is iteratively improved as
       \begin{align}
       \label{condition_max_op}
 \varphi(t)\mapsto \varphi(t) + \epsilon \Im \left(\langle\chi(t)|\left(-\sin\left(\varphi(t)\right)\hat H_1 + \cos\left(\varphi(t)\right)\hat{H}_2\right)|\psi(t)\rangle\right),
\end{align}
where $\epsilon$ is a small positive parameter. The states $|\psi(t)\rangle$ and $|\chi(t)\rangle$ are respectively propagated forward and backward in time from their initial and final conditions. Figure~\ref{oct_BEC} shows three numerical examples of state-to-state transfers. For the three examples, the initial state of the BEC is $|\phi_{0+0}\rangle$. The target states are described in the basis $|\phi_{0+n}\rangle$ with $q=0$. They correspond either to a unique momentum state, a superposition of momentum states or to a Gaussian or a squeezed state~(see~\cite{dupont23} for the mathematical definition of these states). We denote by $|g(x_c,p_,\xi)\rangle$ a squeezed state centered in $(x_c,p_c)$ with the squeezing parameter $\xi$.  The following numerical parameters are used in the numerical simulation: $s=5$, $n_{\textrm{max}}=10$ and $t_f=7.6$ (of the order of $150 \ \mu s$ in real units). More precisely, the target states are then the momentum state $|\phi_{0+2}\rangle$, the superposition $1/\sqrt{3}(|\phi_{0-2}\rangle+|\phi_{0+0}\rangle+|\phi_{0+2}\rangle)$, and the centered squeezed state $|g(x_c=0,p_c=0,\xi=1/3)\rangle$.
\section{Optimal control in the non-linear case}\label{sec3}
Although most experiments to date present a very good match with numerical simulations from the linear model system given in Eq.~(\ref{sysBEC1}), improvement can be made by taking into account the atom interactions which have been neglected in a first step due to the low density of the condensate. This interaction can be modeled at the mean-field level by the Gross–Pitaevskii equation, i.e. by an effective term in the Hamiltonian~(\ref{sysBEC1}), proportional to the square modulus of the wave function $|\psi(x,t)|^2$~\cite{dalfovo1999,leggett2001}. This additional term makes the numerical resolution of the problem more complex. Indeed the non-linear interaction term does not have a simple state-independent matrix representation in the basis $(|\phi_{q+n}\rangle)_{n\in\mathbb{Z}}$.
We present in this section an efficient numerical method which allows rapid propagation of the dynamics, taking into account the interactions between the atoms of the condensate and allows the application of an optimal control algorithm.
\subsection{The model system}
We consider the dimensionless equation~(\ref{sysBEC1}) which describes the dynamics of a BEC in position representation. The interaction between the atoms can be treated at first order as a mean field effect, leading to an additional non-linear term $\lvert \psi \rvert^2$ and to the Gross–Pitaevskii equation,
\begin{equation}
  \label{sysGP}
  \imath \frac{\partial \psi(x,t)}{\partial t} =\left(-\frac{\partial^2}{\partial x^2} - \frac{s}{2}\cos\left( x + \varphi(t) \right) +\beta \lvert \psi(x,t) \rvert^2 \right)\psi(x,t).
\end{equation}
The weight of the non-linear term is given by the dimensionless parameter $\beta$. In the experiments, this parameter is of the order of $0.5\pm0.2$~\cite{tutorial24}.  This value depends on the number of atoms and on the frequency of the hybrid trap.

Expanding the state $|\psi\rangle$ over the eigenvectors of the momentum operator $|\psi\rangle = \sum_{n\in\mathbb{Z}}c_{q,n} |\phi_{q+n}\rangle$, the non-linear term $\lvert \psi \rvert^2$ can be expressed in position representation as,
    \begin{equation*}
  \left\lvert \psi \right\rvert^2 =  \psi(x,t)^*\psi(x,t) = \left(\frac{1}{\sqrt{2\pi}} \sum_{m\in \mathbb{Z}}c_{q,m}^*e^{-\imath (q+m)x}\right) \left( \frac{1}{\sqrt{2\pi}}\sum_{l\in \mathbb{Z}}c_{q,l}e^{\imath (q+l)x}\right)
\end{equation*}
which leads to
$$
\left\lvert \psi \right\rvert^2 = \frac{1}{2\pi}\sum_{m,l \in \mathbb{Z}} c_{q,m}^*c_{q,l} e^{\imath (q+l-m)x}.
$$
The dynamics of the coefficients $c_{q,n}$ are then given by
   \begin{equation}
   \label{lolhGP} \imath \dot{c}_{q,n} = \left( n +q \right)^2c_{q,n} - \frac{s}{4}\left(e^{\imath\varphi(t)}c_{q,n-1} + e^{-\imath\varphi(t)}c_{q,n+1} \right) + \frac{\beta}{2\pi} \sum_{m,l\in\mathbb{Z}} c_{q,m}^*c_{q,l} c_{q,n+m-l}.
\end{equation}
Unlike the Schr\"odinger equation~(\ref{sysBEC}), an analytical matrix representation to this non-linear problem cannot be given here. This means that the simple matrix exponential method used earlier cannot be used anymore to propagate the dynamics. A first option consists in applying the Runge-Kutta of order 4 approach (RK4). However, for large dimensional systems, this method can be very costly in terms of calculation time. Typically a propagation of Eq.~(\ref{lolhGP}) takes about $15$ seconds, for a constant control $\varphi(t)=0$ (see Fig.~\ref{fig1}). Numerical simulations have been performed on a standard laptop. Optimal control, which requires a hundred propagation, is therefore difficult to apply with this approach, as the design of a single control would require a computational time of the order of one hour. In order to speed up the optimization process, we propose in this paper to combine optimal control algorithms with the FBR-DVR bases as described in Sec.~\ref{secDVR}.

\subsection{The FBR-DVR bases}\label{secDVR}
An efficient approach to express the non-linear term of Eq.~(\ref{sysGP}) in matrix form is the FBR-DVR (or pseudo-spectral) approach~\cite{littlejohn2002,light1992,leforestier91,kosloff83,guerin99}. This method consists in using two different bases for the matrix representation of the operators. In our case, the FBR basis (for \emph{Finite Basis Representation}) is the basis of the eigenvectors $|\phi_{q+n}\rangle$ of the momentum operator. In this basis, it is straightforward to express the operators $\hat{p}$ and $\cos(\hat{x})$. The DVR basis (for \emph{Discrete Variable Representation}) corresponds to a basis built from the discretization of a continuous parameter, here the position $x$. In this basis, a matrix representation of the non-linear term $\lvert \psi(x,t) \rvert^2$ can be given. A unitary transformation allows to go from one basis to the other, and therefore to express all the operators in the FBR basis for example. The time propagation of the system can then be done by the matrix exponential approach.

We apply this method to Eq.~\eqref{sysGP}. The FBR basis corresponds to the basis of the eigenvectors $|\phi_{q+n}\rangle$ of the momentum operator. From a numerical point of view, we work in a finite dimensional sub-space $\mathcal{H}_N$ of $\mathcal{H}$ of dimension $N=2n_{\textrm{max}}+1$, with $n_{\textrm{max}} \in \mathbb{N}$ and $-n_{\textrm{max}} \leq n \leq n_{\textrm{max}}$. The potential energy is a periodic spatial function, of period $2\pi$. We choose the interval $[0,2\pi]$ for the position $x$. The DVR basis is built from the discretization of the variable $x \in \left[0, 2\pi\right]$, such that $x_j = \frac{2\pi}{N}j$ with $0 \leq j < N$. A scalar product is given by,
\begin{align*}
    \forall \Psi, \Phi \in \mathcal{H}_N, \ \int_0^{2\pi} \Psi^*(x)\Phi(x)dx = \frac{2\pi}{N}\sum_{j=0}^{N-1}  \Psi^*(x_j)\Phi(x_j).
\end{align*}
From the orthogonality of functions $\phi_{q+n}$ we deduce that we have approximately:
\begin{equation}
    \label{RtR}\frac{2\pi}{N}\sum_{j=0}^{N-1} \phi^*_{q+m}(x_j)\phi_{q+n}(x_j) = \delta_{m,n}.
\end{equation}
From the definition of the plane wave functions, we have also for $n_{\textrm{max}}$ large enough:
\begin{equation}
   \label{RRt} \frac{2\pi}{N}\sum_{n=-n_{\textrm{max}}}^{n_{\textrm{max}}} \phi_{q+n}(x_i)\phi_{q+n}^*(x_j) = \delta_{i,j}.
\end{equation}
We introduce the matrix $\hat{R}$ defined as
       \begin{align}
       \centering
  \label{R_matrix} \hat{R} =  \begin{pmatrix}
    R_{0,-n_{\textrm{max}}} & R_{0,-n_{\textrm{max}}+1} & \dots \\
    \vdots & \ddots & \\
    R_{N-1,-n_{\textrm{max}}} &        & R_{N-1,n_{\textrm{max}}}
  \end{pmatrix},
\end{align}
with matrix elements  $R_{j,n}$ given by
\begin{align}
   R_{j,n} = \frac{1}{\sqrt{N}}e^{\imath \frac{2\pi}{N}\left(q + n\right)j}.
\end{align}
Equations \eqref{RtR} and \eqref{RRt} are equivalent to $\hat{R}^{\dagger}\hat{R}=\hat{I}$ and $\hat{R}\hat{R}^{\dagger}=\hat{I}$. The matrix $\hat{R}$ is thus a unitary matrix. The DVR basis $\left(|u_j\rangle\right)_{j\in[0,N-1]}$ is then built from the following kets as
\begin{equation}
   |u_j\rangle = \sum_{n=n_{\textrm{max}}}^{-n_{\textrm{max}}}\hat{R}_{jn}^*|\phi_n\rangle.
\end{equation}
This basis connects a basis of $\mathcal{H}_N$ to the spatial discretization basis. We deduce that an operator $\hat{W}$ depending on $x$ has a diagonal representation in the DVR basis given by
\begin{equation}
   \langle u_i|\hat{W}|u_j\rangle \simeq W(x_j) \delta_{i,j}.
\end{equation}
The matrix $\hat{R}$ allows to express a known operator in the DVR basis to the FBR basis, and vice versa
\begin{equation}
   \hat{W}^{\textrm{FBR}} = \hat{R}^{\dagger} \hat{W}^{\textrm{DVR}}\hat{R}.
\end{equation}
Since the square modulus of the wave function, $\lvert \psi(x) \rvert^2$, depends on the position $x$ as shown in Eq~\eqref{lolhGP}, it is straightforward to obtain its matrix representation in the DVR basis, and then in the FBR basis. The exponential matrix approach can then be used to propagate the dynamics. We denote by $\hat{G}^{\textrm{DVR}}$ the matrix of $\lvert \psi(x) \rvert^2$ in the DVR basis,
       \begin{align}
       \centering
  \label{G_matrix} \hat{G}^{\textrm{DVR}} =  \begin{pmatrix}
   G_0 &  &  & \\
   & G_{1} &  & \\
   &  &  \ddots & \\
   &  &   & G_{N-1}
  \end{pmatrix},
\end{align}
where
\begin{align}
   G_j = \displaystyle\left\lvert  \frac{1}{\sqrt{2\pi}} \sum_{n=-n_{\textrm{max}}}^{n_{\textrm{max}}} c_{q,n} e^{\imath \frac{2\pi}{N}(q+n)j} \right\rvert^2.
\end{align}
The parameters $c_{q,n}$ are the coefficients of the state in the FBR basis.
Using the notations introduced in Sec.~\ref{sec2} and the relations~\eqref{R_matrix} and~\eqref{G_matrix}, Eq.~\eqref{sysGP} can be written in matrix form in the FBR basis as follows
    \begin{equation}
  \label{dynEtaNL} \imath \frac{d |\psi\rangle }{dt} = \hat{H}_{GP} |\psi\rangle = \left( \hat{H}_0 + \cos(\varphi(t)) \hat{H}_1 + \sin(\varphi(t))\hat{H}_2 + \beta \hat{R}^{\dagger} \hat{G}^{DVR}\hat{R} \right) |\psi\rangle,
\end{equation}
with $\hat{H}_{\textrm{GP}}$ the Gross–Pitaevskii Hamiltonian.

\begin{figure}[h!]
\centering
\includegraphics[width=0.5\textwidth]{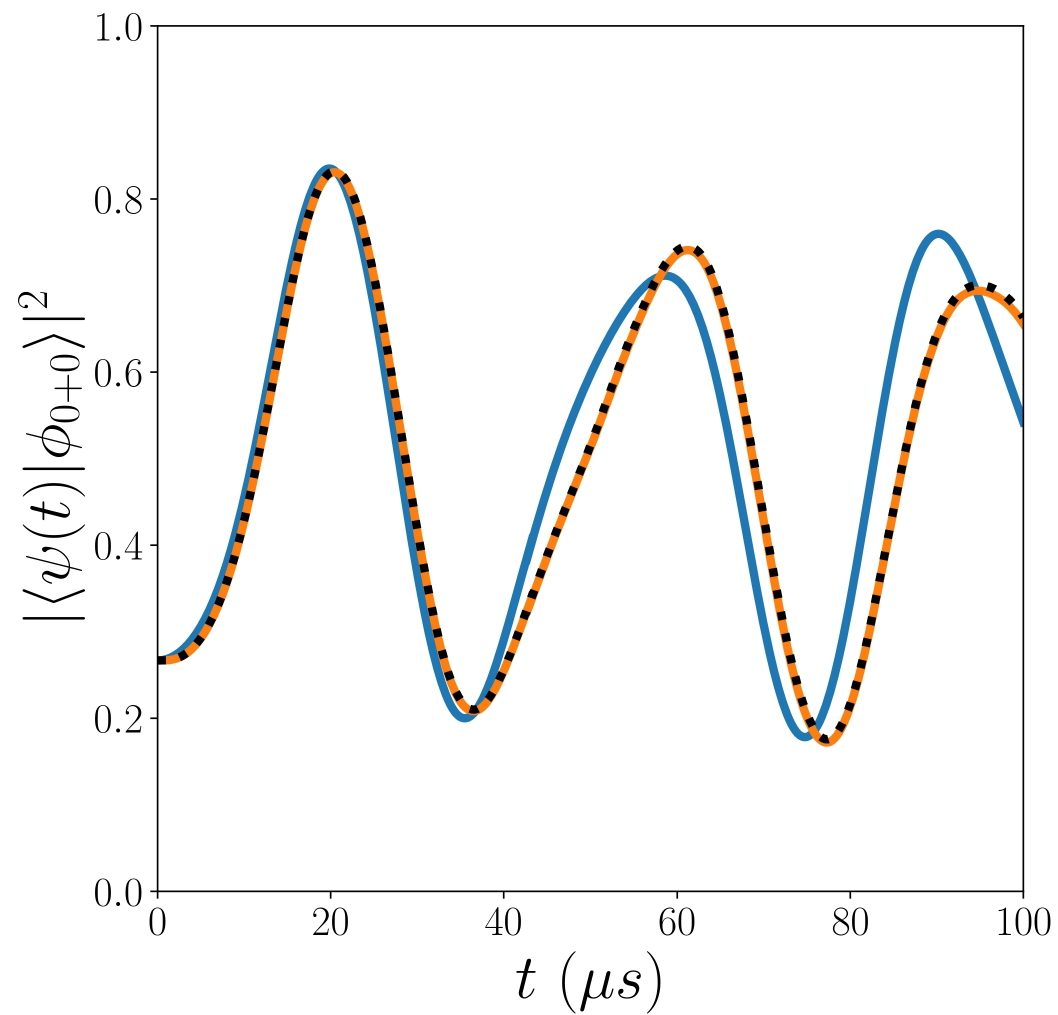}
\caption{Time evolution of $|\langle\psi(t)|\phi_{0+0}\rangle |^2$ for a zero control, $\varphi=0$. The initial state is $|\psi(0)\rangle = |g(0,0,1/2)\rangle$. The blue and orange solid lines represent respectively the propagation of the state with the matrix exponential approach for $\beta=0$ and $\beta=1$. The dashed black line corresponds to a dynamic computed with the RK4 method for $\beta=1$.}
\label{fig1}
\end{figure}

For comparison, the propagation for a constant control is here of the order of $0.6$ seconds, i.e. a total of around 2 minutes for the calculation time of the optimal control. Figure~\ref{fig1} shows the time evolution of the projection of the state $|\psi\rangle$ on  $|\phi_{0+0}\rangle$, calculated from the matrix exponential approach and the RK4 method. The initial state is the squeezed state  $|g(0,0,1/2)\rangle$ and the phase $\varphi(t)=0$. For $\beta=1$, we observe that the two methods give the same result.


\subsection{The non-linear GRAPE}
We show in this section how to extend the standard GRAPE algorithm to this non-linear case. The corresponding algorithm is called \emph{non-linear GRAPE}.

We consider the state-to-state transfer defined with the fidelity $F=|\langle\psi_t|\psi(t_f)\rangle|^2$, with the dynamics of Eq.~\eqref{sysGP}.
In this case, the Pontryagin Hamiltonian $H_p$ can be written as
    \begin{equation}
    H_p = \Im [\langle\chi|\hat{H}_{GP}|\psi\rangle],
\end{equation}
with $\hat{H}_{GP}$, the Gross–Pitaevskii Hamiltonian in position representation,
    \begin{equation}
     \hat{H}_{GP} =  -\frac{\partial^2}{\partial x^2} - \frac{s}{2}\cos\left( x + \varphi(t) \right) + \beta \lvert \psi(x,t) \rvert^2.
\end{equation}
The time evolution of the adjoint state is given by the relation, $\dot{\chi}(x,t)=-\partial_{\psi}H_p$, which leads to
    \begin{equation}
   \label{etat_p} \imath\frac{\partial \chi}{\partial t} = \left( -\frac{\partial^2}{\partial x^2} - \frac{s}{2}\cos\left(x + \varphi(t)\right) + 2\beta \displaystyle\left\lvert \psi \right\rvert^2 \right) \chi - \beta \psi^2 \chi^*.
\end{equation}
Note that the nonlinearity of the Gross-Pitaeski equation breaks the symmetry between the state and the adjoint state which are no longer solutions of the same differential equation. The transversality condition on the adjoint state $\chi(x,t_f) = \chi_0 \partial_{\psi} F$ yields
    \begin{equation}
    \chi(x,t_f) = 2\chi_0 \langle\psi_t|\psi(t_f)\rangle\psi_t,
\end{equation}
with $\chi_0=1/2$ in the numerical simulations. Finally, the maximization condition of the Pontryagin Hamiltonian leads to the iterative procedure of GRAPE
$$
\varphi(t)\mapsto \varphi(t)+\epsilon \Im [\langle\chi|\frac{\partial \hat{H}_{\textrm{GP}}}{\partial \varphi}|\psi\rangle ]
$$
where $\epsilon$ is a small positive parameter. The state and the adjoint state are respectively propagated forward and backward in time. The differential equation~\eqref{etat_p} associated with the adjoint state can be written as
    \begin{equation}
   \label{etat_p2} \imath\frac{\partial \chi}{\partial t} = \left( -\frac{\partial^2}{\partial x^2} - \frac{s}{2}\cos(x+\varphi(t)) + \beta \displaystyle\left\lvert \psi \right\rvert^2 \right) \chi - 2\imath\beta \Im\left[\chi^*\psi\right] \psi.
\end{equation}
The backward propagation can be done in matrix form with the final conditions $|\psi(t_f)\rangle = |\psi_f\rangle$ and $|\chi(t_f)\rangle = \langle\psi_t|\psi(\tau_f)\rangle |\psi_t\rangle$, with the following extended system
    \begin{equation}
    \imath\begin{pmatrix}
|\dot{\psi}\rangle \\ |\dot{\chi}\rangle
    \end{pmatrix} =     \begin{pmatrix}
\hat{H}_{\textrm{GP}} & 0 \\
  - 2\imath\beta \hat{R}^{\dagger}\hat{I}^{\textrm{DVR}}\hat{R} & \hat{H}_{\textrm{GP}}
    \end{pmatrix}\begin{pmatrix}
|\psi\rangle \\ |\chi\rangle
    \end{pmatrix},
\end{equation}
where $\hat{H}_{\textrm{GP}}$ is the Gross–Pitaevskii Hamiltonian in matrix form given by Eq.~\eqref{dynEtaNL}, and $\hat{I}^{\textrm{DVR}}$ is the diagonal matrix associated with the term $\Im\left[\chi^*\psi\right]$ in the DVR basis. The matrix elements of $\hat{I}^{\textrm{DVR}}$ are $I_j = \Im\left[ \frac{1}{2\pi}  \sum_{m,n} d_{q,m}^*c_{q,n} e^{\imath \frac{2\pi}{N}(q+n-m)j} \right]$, $d_{q,n}$ being the coefficients of the adjoint state $|\chi\rangle$ in the FBR basis, $-n_{\textrm{max}}\leq n,m\leq n_{\textrm{max}}$.

\begin{figure}
    \centering
        \includegraphics[width=0.5\textwidth]{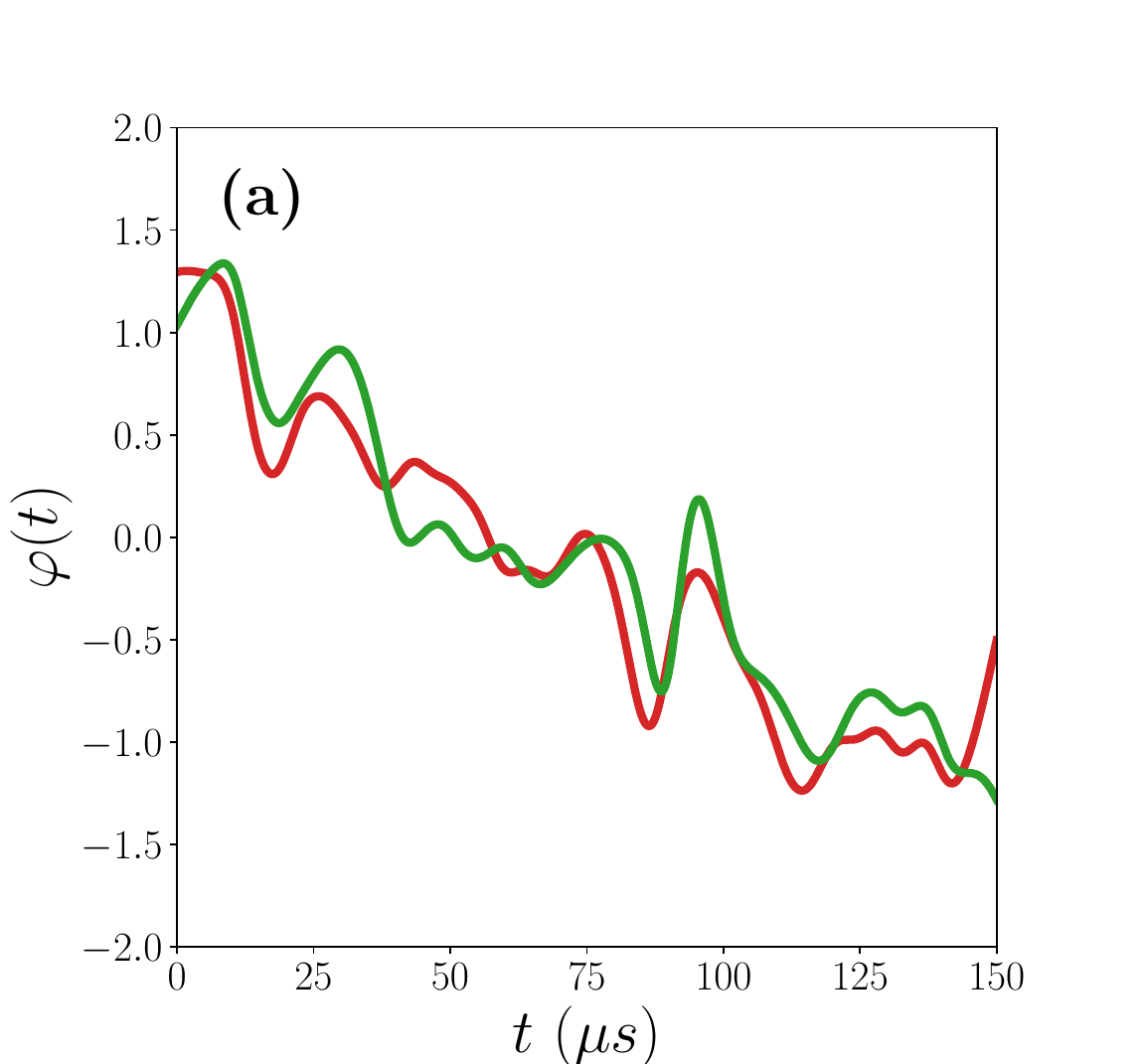}
        \includegraphics[width=0.5\textwidth]{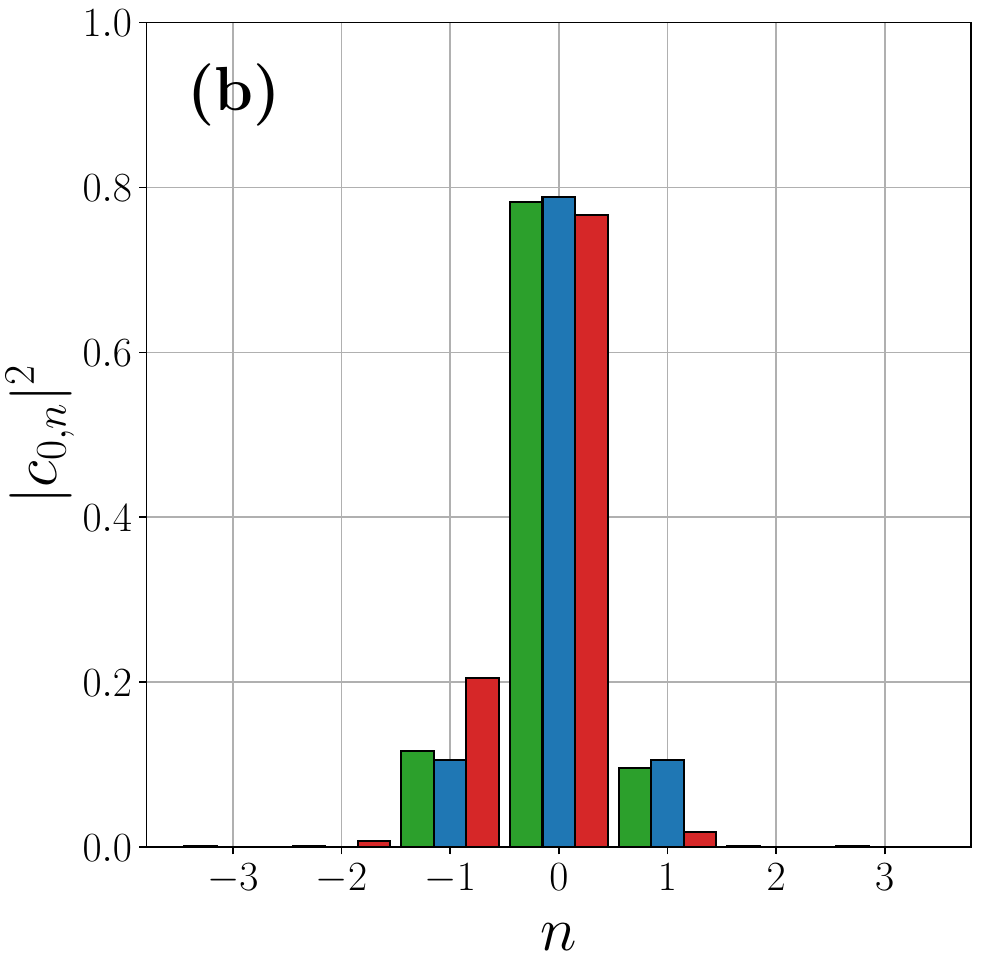}
        \caption{(a)) Time evolution of the optimal controls for $\beta=0$ (red line) and $\beta=0.7$ (green line). (b) Populations $|c_{0,n}|^2$ at the final time. The blue, green and red columns correspond respectively to the populations of the target state $|g(0,0,3/2)\rangle$, of the final state for a control optimized with $\beta=0.7$ and of the final state for a control with $\beta=0$.}
        \label{fignonlinear}
\end{figure}

\begin{figure}
    \centering
        \centering
        \includegraphics[width=0.5\textwidth]{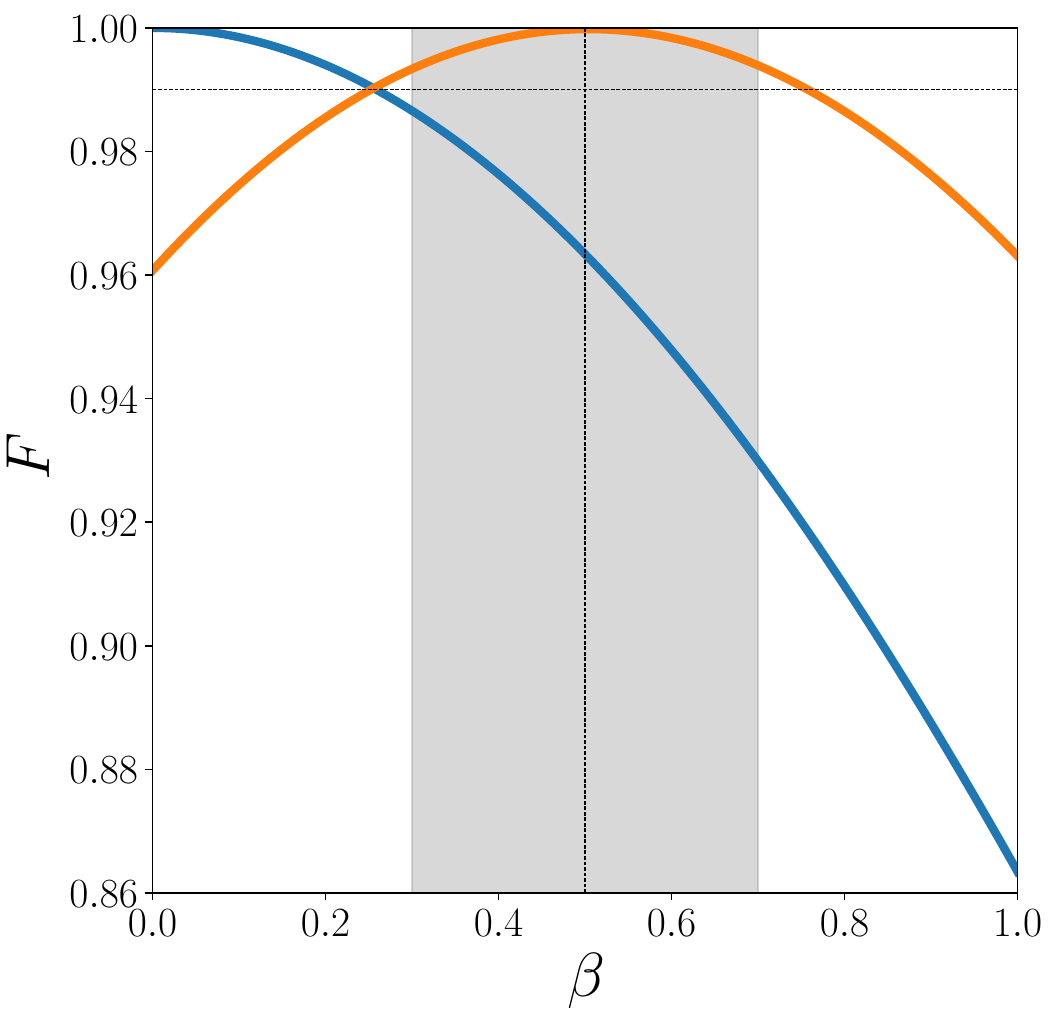}
        \caption{Evolution of the fidelity $F$ for the target state $|g(0,0,3/2)\rangle$ as a function of $\beta$. The blue and orange lines correspond respectively to $\beta=0$ and $\beta=0.5$.} 
        \label{fignonlinear2}
\end{figure}

As an illustrative example, we consider the transfer from the state $|\psi(0)\rangle = |\phi_{0+0}\rangle$ to the squeezed state $|g\left(0,0,3/2\right)\rangle$. The control time is set to $t_f = 150 ~\mu s$, the lattice depth to $s=5$, and the quasi-momentum to $q=0$. A first control is optimized for $\beta=0$ as represented in Fig.~\ref{fignonlinear}. The dynamics are then propagated for this control with a non-linear coefficient $\beta$ going from $0$ to $1$, $\beta \in [0,1]$. Figure~\ref{fignonlinear2} shows the evolution of the fidelity with respect to the nonlinear parameter $\beta$. The grey area indicates the experimental uncertainty on this parameter. We observe that for $\beta=0.7$, the upper bound of the uncertainty interval, the fidelity is only of the order of 0.93, which means that the target state is not reached by the system. A second control is optimized by using the non-linear version of GRAPE for $\beta=0.5$. In this case, the fidelity is equal to 0.994 for the same control time. Figure~\ref{fignonlinear} depicts the different populations in the momentum eigenbasis. The role of $\beta$ on the dynamics depends on the initial and target states considered for the optimal control.

\section{Optimal control of a BEC in a two-dimensional optical lattice}\label{sec4}
In this section we show how to apply the GRAPE algorithm to a two-dimensional optical lattice. We assume here that the non-linearity of the Gross-Pitaevski equation can be neglected.

Before presenting the modeling of an experiment with a 2D lattice, it may be useful to recall the concept of direct and reciprocal lattices, as well as Bloch's theorem in the multidimensional case.
The direct lattice is a set of vectors describing the nodes of the lattice, i.e. here the positions where the potential is minimum. We denote by $\mathbf{a}_i$ the vectors of the basis of the direct lattice and $N$ the corresponding dimension. A node is an element of this set,
      \begin{equation}
   \label{rdirect} \mathcal{D} = \left\{ \sum_{i=1}^{N} j_i \mathbf{a}_i, \ j_i \in \mathbb{Z} \right\}.
\end{equation}
The reciprocal lattice is described by the set,
      \begin{equation}
   \label{rdeci} \mathcal{R} = \left\{ \sum_{i=1}^{N} j_i \mathbf{b}_i, \ j_i \in \mathbb{Z} \right\}
\end{equation}
where the vectors $\mathbf{b}_i$ are defined from the relation,
      \begin{equation}
   \label{rdeci2} \mathbf{a}_i\cdot \mathbf{b}_j = 2\pi\delta_{ij},
\end{equation}
with $\mathbf{a}_i \in \mathcal{D}$ and $\mathbf{b}_j \in \mathcal{R}$.
Using the Bloch's theorem, it can be shown that the wave function $\psi(\mathbf{r})$ solution of the time-independent Schr\"odinger equation,
      \begin{equation}
   \label{schro_multi}  \left( -\frac{\hbar^2}{2m}\frac{\partial^2}{\partial \mathbf{r}^2} + V(\mathbf{r}) \right) \psi(\mathbf{r})=E\psi(\mathbf{r}),
\end{equation}
where the potential $V$ is periodic, $V(\mathbf{r} + \mathbf{R}) = V(\mathbf{r})$ for any $\mathbf{R} \in \mathcal{D}$ and $\mathbf{r} = \sum_{i=1}^N r_i \mathbf{a}_i$ is the position of the system, can be expressed as
      \begin{equation}
  \label{bloch2D} \psi_{\mathbf{q}}(\mathbf{r}) = e^{\imath \mathbf{q} \cdot \mathbf{r}} u_{\mathbf{q}}(\mathbf{r}),
\end{equation}
with $\mathbf{q}=\sum_{i=1}^N q_i \mathbf{b}_i$ the two-dimensional quasi-momentum and $u_{\mathbf{q}}(\mathbf{r})$ a periodic function in $\mathcal{D}$.
\subsection{The model system}
A one-dimensional optical lattice can be generated by the superposition of two lasers. A two-dimensional optical lattice requires three or more lasers. A large variety of two-dimensional configurations can be realized experimentally~\cite{morsch2006}.
As an illustrative example, we consider the following configuration, three lasers with the same angular frequency $\omega$ which propagate in the plane $\left(\mathbf{e}_x, \mathbf{e}_y \right)$ of the laboratory frame with a linear polarization along the $z$- axis.
The wave vectors are defined in this plane by
      \begin{align*}
   \label{k} \mathbf{k}_1 &= k(1,0) \\
   \mathbf{k}_2 &= \frac{k}{2}(-1,\sqrt{3}) \\
   \mathbf{k}_3 &= -\frac{k}{2}(1,\sqrt{3}),
\end{align*}
where $k=\frac{\omega}{c}$ is the wave number. The angle between the vectors is $2\pi/3$. The total electric field is then given by
  \begin{equation}
   \label{elec} \mathbf{E} = E_0 \sum_i \cos \left( \mathbf{k}_i \cdot \mathbf{r} - \omega t + \varphi_i \right) \mathbf{e}_z,
\end{equation}
where $E_0$ is the amplitude of the electric field, $\mathbf{r}$ the position vector and $\varphi_i$ is the phase of the laser $i$, with $i=1,2,3$. The interaction between the electric field and the atom is described by the Hamiltonian $\hat{H}_I = -\mathbf{\mu}\cdot\mathbf{E}$, where $\mathbf{\mu}$ is the induced dipole moment. At first order, the dipole moment is given by $\mathbf{\mu} = -\frac{\alpha}{2}\mathbf{E}$, where $\alpha$ is the polarizability of the atom. This leads to $\hat{H}_I = -\frac{\alpha}{2}\mathbf{E}\cdot \mathbf{E}$.
The electric field is not resonant with the transition frequencies of the atom, and the period associated with the $\omega$ frequency is much shorter than the characteristic time of the experiment, $T_{\textrm{dyn}}$. The interaction Hamiltonian can be averaged to only consider the long-time dynamics. The BEC is then subjected to the following dipole potential
\begin{align}
   \label{12becz} V = \langle \hat{H}_I \rangle &= -\frac{3}{4}\alpha E_0^2 - \frac{\alpha}{2}E_0^2   \begin{aligned}[t] &\left[ \cos  \left( (\mathbf{k}_1 - \mathbf{k}_2) \cdot \mathbf{r} + \varphi_{12} \right)\right. \\
   & + \cos \left( (\mathbf{k}_2 - \mathbf{k}_3) \cdot \mathbf{r} + \varphi_{23} \right) \\
   & \left. + \cos \left( (\mathbf{k}_3 - \mathbf{k}_1) \cdot \mathbf{r} + \varphi_{31} \right)   \right],
   \end{aligned}
\end{align}
where $\varphi_{jk} = \varphi_j - \varphi_k$, and $\langle \cdot \rangle$ denotes the time average over one period $\tau$, such that $\frac{2\pi}{\omega} \gg \tau \gg T_{\textrm{dyn}}$.
Finally, we deduce that this average potential governs the 2D-  dynamics~(\ref{schro_multi}), with $\mathbf{r} = \left( x,y \right)$.

The direct lattice nodes correspond to the minima of the potential energy for $\varphi_{jk}=0$. We have
      \begin{align}
   \label{neoud} (\mathbf{k}_1 - \mathbf{k}_2) \cdot \mathbf{r} &= 0 \ (\textrm{mod}~2\pi) \nonumber \\
   (\mathbf{k}_2 - \mathbf{k}_3) \cdot \mathbf{r} &= 0 \ (\textrm{mod}~2\pi) \\
   (\mathbf{k}_3 - \mathbf{k}_1) \cdot \mathbf{r} &= 0 \ (\textrm{mod}~2\pi) \nonumber
\end{align}
A node of the direct lattice belongs to the set $\mathcal{D}$~(\ref{rdirect}) with,
      \begin{align*}
    \mathbf{a}_1 &= -\frac{4\pi}{3 k^2} \mathbf{k}_3, \\
   \mathbf{a}_2 &= \frac{4\pi}{3 k^2} \mathbf{k}_2.
\end{align*}
A node of the reciprocal lattice is an element of the set $\mathcal{R}$~(\ref{rdeci}) with
      \begin{align*}
    \mathbf{b}_1 &= \mathbf{k}_1 - \mathbf{k}_3, \\
   \mathbf{b}_2 &= \mathbf{k}_2 - \mathbf{k}_1.
\end{align*}
Since the potential is periodic, Bloch's theorem applies, and eigenvectors are as in Eq.~(\ref{bloch2D}). Within a given sub-Hilbert space of quasi-momentum $\mathbf{q}$, a generic wave function can be written as:
    \begin{align}
   \label{2intbis} \psi(\mathbf{r}) = \sum_{m,n \in \mathbb{Z}} c_{\mathbf{q}, m,n} e^{\imath \mathbf{q}\cdot \mathbf{r}} e^{\imath\left( m\mathbf{b}_1 + n\mathbf{b}_2 \right)\cdot \mathbf{r}}.
\end{align}
We denote by $\phi_{m,n}=e^{\imath \left( m\mathbf{b}_1 + n\mathbf{b}_2 \right)\cdot \mathbf{r}}$ the functions of the Fourier basis. The coefficients $c_{\mathbf{q},m,n}$ are the coefficients of the state in this basis.
In order to simplify the study, we consider below that the quasi-momentum is zero, which is a reasonable experimental assumption~\cite{dupont21}. Plugging the expression of $\psi(\mathbf{r},t)$ of Eq.~\eqref{2intbis} into Eq.~\eqref{schro_multi}, we obtain,
 \begin{align*}
   \label{3int} \imath\hbar\sum_{m,n \in \mathbb{Z}} \dot{c}_{m,n} \begin{aligned}[t]& e^{\imath\left( m\mathbf{b}_1 + n\mathbf{b}_2 \right)\mathbf{r}} =
   \sum_{m,n \in \mathbb{Z}} \frac{\hbar^2}{2m} c_{m,n} (m\mathbf{b}_1 + n\mathbf{b}_2)^2 e^{\imath\left( m\mathbf{b}_1 + n\mathbf{b}_2 \right)\mathbf{r}} \\
   &- \sum_{m,n \in \mathbb{Z}} \frac{V_0}{4} c_{m,n} \left[ e^{\imath\left((\mathbf{k}_1-\mathbf{k}_2)\cdot \mathbf{r} + \varphi_{12}\right)} e^{\imath\left( m\mathbf{b}_1 + n\mathbf{b}_2 \right)\mathbf{r}} + e^{-\imath\left((\mathbf{k}_1-\mathbf{k}_2)\cdot \mathbf{r} + \varphi_{12}\right)} e^{\imath\left( m\mathbf{b}_1 + n\mathbf{b}_2 \right)\mathbf{r}} \right] \\
   &- \sum_{m,n \in \mathbb{Z}} \frac{V_0}{4} c_{m,n} \left[ e^{\imath\left((\mathbf{k}_2-\mathbf{k}_3)\cdot \mathbf{r} + \varphi_{23}\right)} e^{\imath\left( m\mathbf{b}_1 + n\mathbf{b}_2 \right)\mathbf{r}} + e^{-\imath\left((\mathbf{k}_2-\mathbf{k}_3)\cdot \mathbf{r} + \varphi_{23}\right)} e^{\imath\left( m\mathbf{b}_1 + n\mathbf{b}_2 \right)\mathbf{r}} \right] \\
   &- \sum_{m,n \in \mathbb{Z}} \frac{V_0}{4} c_{m,n} \left[ e^{\imath\left((\mathbf{k}_3-\mathbf{k}_1)\cdot \mathbf{r} + \varphi_{31}\right)} e^{\imath\left( m\mathbf{b}_1 + n\mathbf{b}_2 \right)\mathbf{r}} + e^{-\imath\left((\mathbf{k}_3-\mathbf{k}_1)\cdot \mathbf{r} + \varphi_{31}\right)} e^{\imath\left( m\mathbf{b}_1 + n\mathbf{b}_2 \right)\mathbf{r}} \right],
   \end{aligned}
\end{align*}
where the constant term $-\frac{3}{4}\alpha E_0^2$ of the potential $V$ is removed, and $V_0=\alpha E_0^2$ is the lattice depth. From the relations, $\mathbf{k}_1 - \mathbf{k}_2  = - \mathbf{b}_2$, $\mathbf{k}_2 - \mathbf{k}_3  = \mathbf{b}_1 + \mathbf{b}_2$, $\mathbf{k}_3 - \mathbf{k}_1  = - \mathbf{b}_1$, $\mathbf{b}_1 \cdot \mathbf{b}_1 = \mathbf{b}_2 \cdot \mathbf{b}_2 = 3 k^2$ and $\mathbf{b}_1 \cdot \mathbf{b}_2 = -\frac{3}{2}k^2$, we arrive at
\begin{align}
   \imath\hbar\sum_{m,n \in \mathbb{Z}} \Dot{c}_{m,n} 
   & e^{\imath\left( m\mathbf{b}_1 + n\mathbf{b}_2 \right)\mathbf{r}} = \sum_{m,n \in \mathbb{Z}} \frac{3k^2\hbar^2}{2m} c_{m,n} (m^2 + n^2 - mn) e^{\imath\left( m\mathbf{b}_1 + n\mathbf{b}_2 \right)\mathbf{r}} \nonumber \\
   &- \sum_{m,n \in \mathbb{Z}} \frac{V_0}{4} c_{m,n} \left[ e^{\imath\varphi_{12}} e^{\imath\left( m\mathbf{b}_1 + (n-1)\mathbf{b}_2 \right)\mathbf{r}} + e^{-\imath\varphi_{12}} e^{\imath\left( m\mathbf{b}_1 + (n+1)\mathbf{b}_2 \right)\mathbf{r}} \right] \nonumber \\
   &- \sum_{m,n \in \mathbb{Z}} \frac{V_0}{4} c_{m,n} \left[ e^{\imath\varphi_{23}} e^{\imath\left( (m+1)\mathbf{b}_1 + (n+1)\mathbf{b}_2 \right)\mathbf{r}} + e^{-\imath\varphi_{23}} e^{\imath\left( (m-1)\mathbf{b}_1 + (n-1)\mathbf{b}_2 \right)\mathbf{r}} \right] \nonumber \\
   &- \sum_{m,n \in \mathbb{Z}} \frac{V_0}{4} c_{m,n} \left[ e^{\imath \varphi_{31}} e^{\imath\left( (m-1)\mathbf{b}_1 + n\mathbf{b}_2 \right)\mathbf{r}} + e^{-\imath \varphi_{31}} e^{\imath\left( (m+1)\mathbf{b}_1 + n\mathbf{b}_2 \right)\mathbf{r}} \right].\nonumber
\end{align}
Projecting this equation onto the state $\phi_{m,n}$ and introducing the dimensionless time $\tau = \frac{3 \hbar k^2}{2m}t$ and lattice depth $s=2mV_0/(3\hbar^2k^2)$, we obtain the differential equation governing the dynamics of the coefficients $c_{m,n}$,
\begin{eqnarray}
   \label{coeff2D} & \imath\dot{c}_{m,n} = (m^2 + n^2 - mn)c_{m,n}
   - \frac{s}{4} \left[ e^{\imath\varphi_{12}} c_{m,n+1} + e^{-\imath\varphi_{12}} c_{m,n-1} \right. \\
  &  + e^{\imath\varphi_{23}} c_{m-1,n-1} + e^{-\imath\varphi_{23}} c_{m+1,n+1}+\left. e^{\imath\varphi_{31}} c_{m+1,n} + e^{-\imath\varphi_{31}} c_{m-1,n} \right].\nonumber
   \end{eqnarray}

\subsection{Optimal control problem}

At this point, Eq.~\eqref{coeff2D} can be put in matrix from. We have
\begin{equation*}
\imath \frac{d|\psi(t)\rangle}{dt} =\left(\hat{H}_0 + e^{\imath \varphi_{12}}\hat{H}_{12}^+ + e^{-\imath \varphi_{12}}\hat{H}_{12}^- + e^{\imath \varphi_{23}}\hat{H}_{23}^+ + e^{-\imath \varphi_{23}}\hat{H}_{23}^- + e^{\imath \varphi_{31}}\hat{H}_{31}^+ + e^{-\imath \varphi_{31}}\hat{H}_{31}^-\right)|\psi(t)\rangle,
\end{equation*}
with
$$
\begin{cases}
\hat{H}_0 = \sum_{m,n\in\mathbb{Z}} \left( m^2 + n^2 - mn \right) |\phi_{m,n}\rangle\langle\phi_{m,n}| \\
\hat{H}_{12}^+ = -\frac{s}{4}\sum_{m,n\in\mathbb{Z}} |\phi_{m,n}\rangle\langle\phi_{m,n+1}| \\
\hat{H}_{12}^- = -\frac{s}{4}\sum_{m,n\in\mathbb{Z}} |\phi_{m,n}\rangle\langle\phi_{m,n-1}| \\
\hat{H}_{23}^+ = -\frac{s}{4}\sum_{m,n\in\mathbb{Z}} |\phi_{m,n}\rangle\langle\phi_{m-1,n-1}| \\
\hat{H}_{23}^- = -\frac{s}{4}\sum_{m,n\in\mathbb{Z}} |\phi_{m,n}\rangle\langle\phi_{m+1,n+1}| \\
\hat{H}_{31}^+ = -\frac{s}{4}\sum_{m,n\in\mathbb{Z}} |\phi_{m,n}\rangle\langle\phi_{m+1,n}| \\
\hat{H}_{31}^- = -\frac{s}{4}\sum_{m,n\in\mathbb{Z}} |\phi_{m,n}\rangle\langle\phi_{m-1,n}|
\end{cases}
$$
From a numerical point of view, we use a Hilbert space of finite dimension such that $-M\leq m \leq M$ and $-N\leq n \leq N$. The dimension of this space is  $d_{\mathcal{H}}=(2M+1)\times (2N+1)$. In the numerical simulations, $M$ and $N$ have been set to 5. The GRAPE algorithm can then be applied to this case~\cite{tutorial24}. For each control $\varphi_{12}$, $\varphi_{23}$ and $\varphi_{31}$, the maximization condition given by the PMP leads to the iterative procedure of GRAPE
\begin{equation*}
 \varphi_\ell\mapsto \varphi_\ell+\epsilon     \Im \left[ \langle\chi(t)| \imath \left(\hat{H}_{\ell}^+ - \hat{H}_{\ell}^- \right) |\psi(t)\rangle \right],
\end{equation*}
where $|\chi\rangle$ is the adjoint state and $\ell$ corresponds to the index $12$, $23$ or $31$.

\begin{figure}
    \centering
     \includegraphics[width=0.5\textwidth]{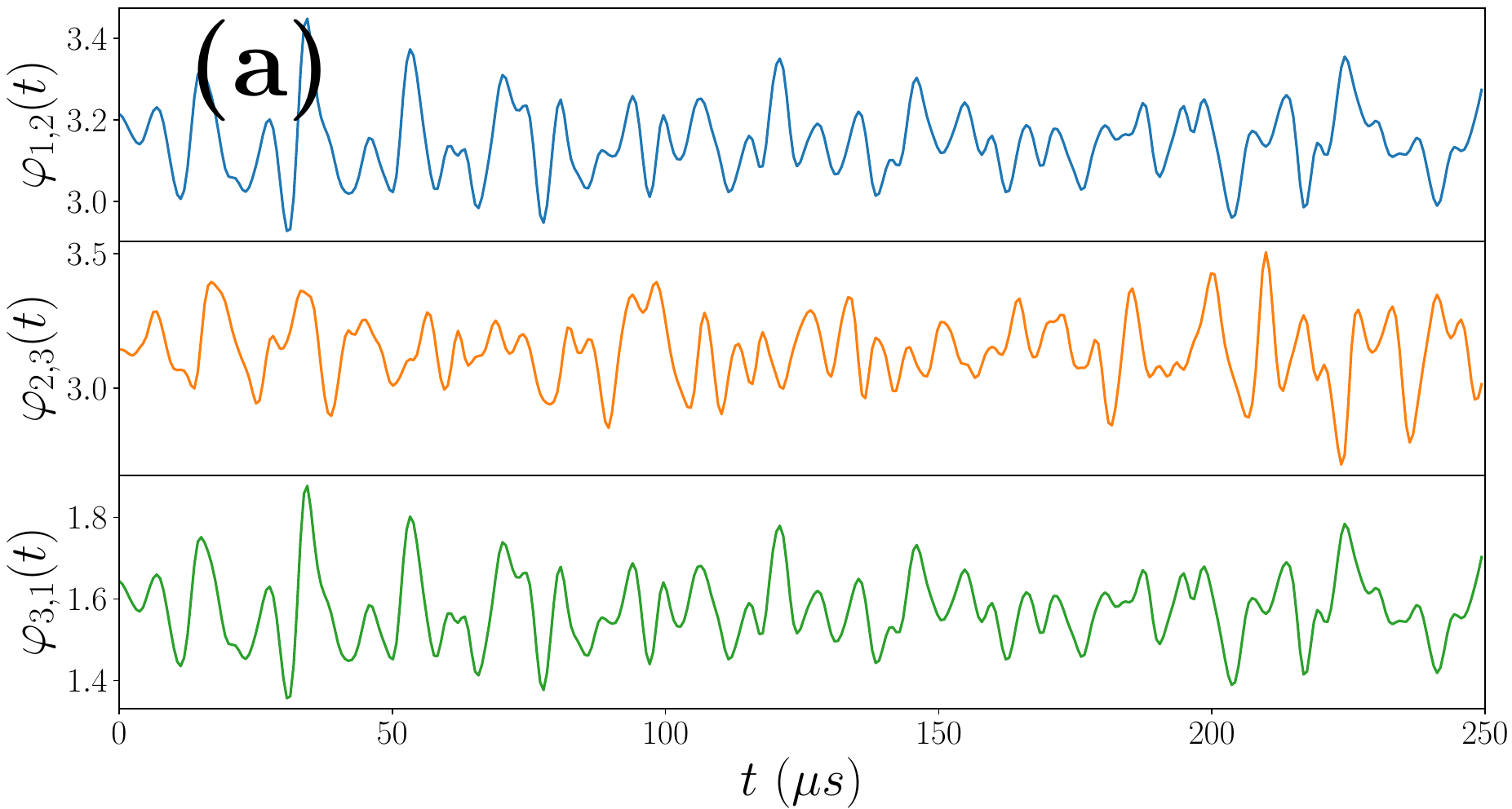}
     \includegraphics[width=0.5\textwidth]{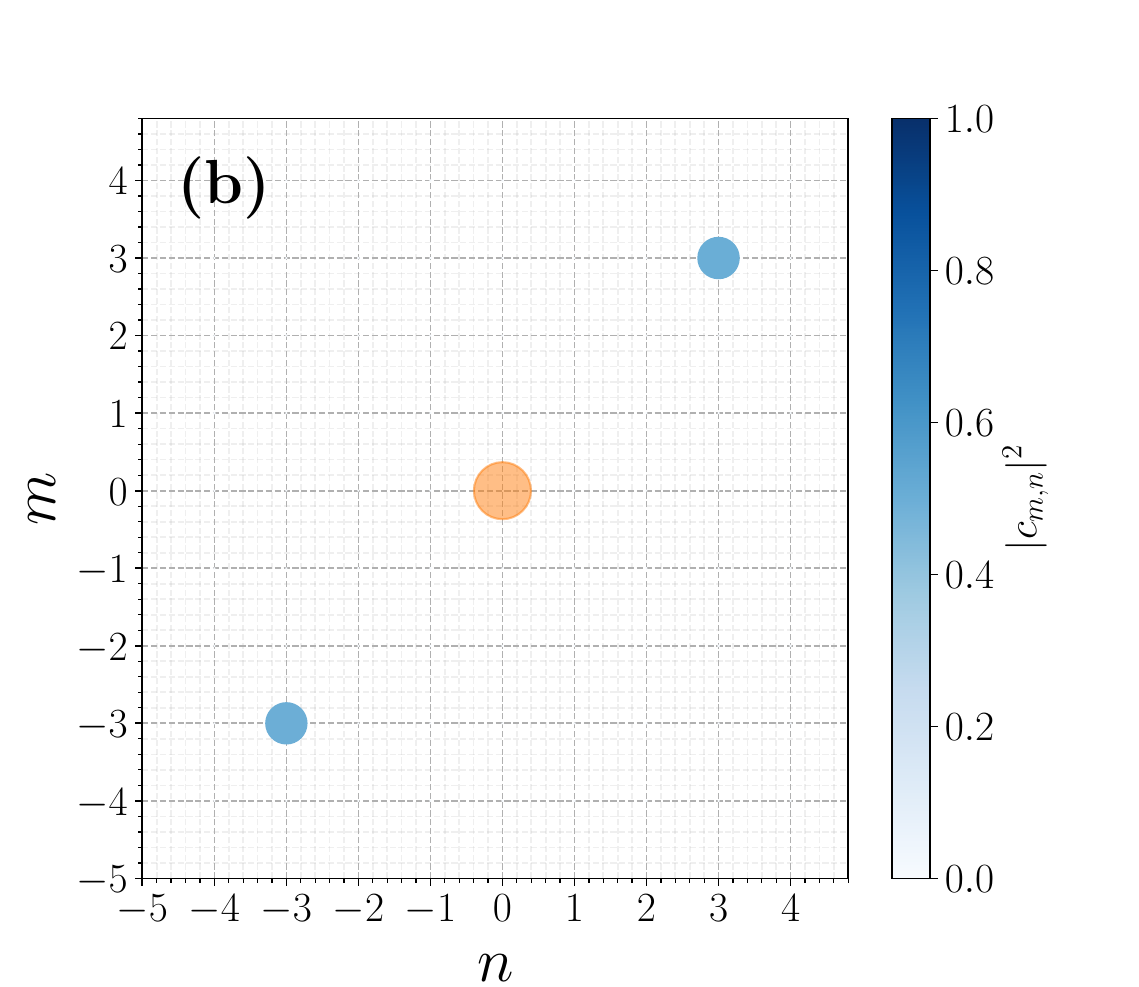}
   \caption{Transfer from the state $|\phi_{0,0}\rangle$ to the target $\frac{1}{\sqrt{2}}\left( |\phi_{-3,-3}\rangle + |\phi_{3,3}\rangle \right)$; (a) Time evolution of the controls $\varphi_{12}$, $\varphi_{23}$ and $\varphi_{31}$, (b) Populations of the initial (orange) and target (blue) states.}
    \label{fig:2Dpopcont}
\end{figure}

\begin{figure}
    \centering
     \includegraphics[width=0.5\textwidth]{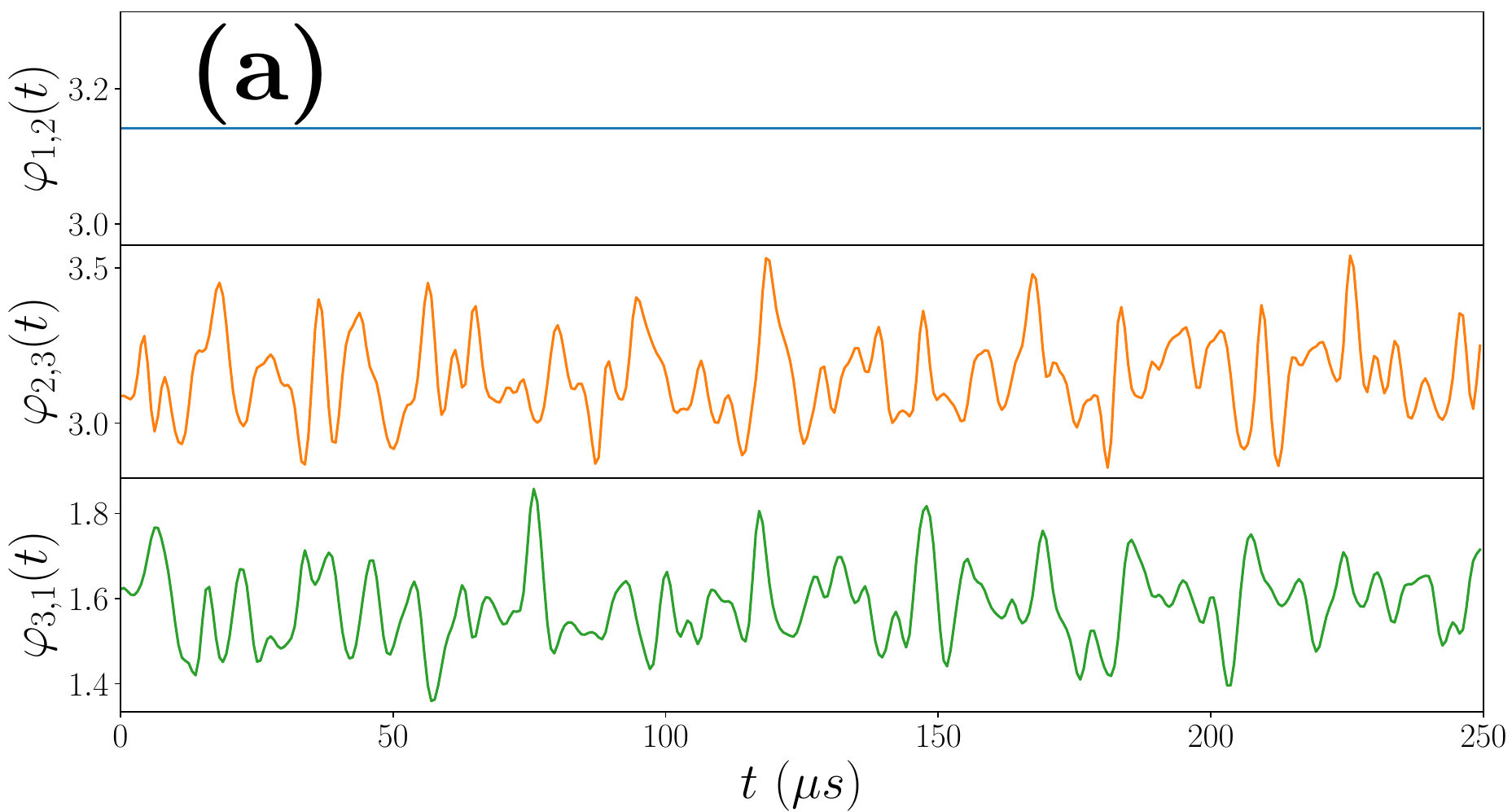}
     \includegraphics[width=0.5\textwidth]{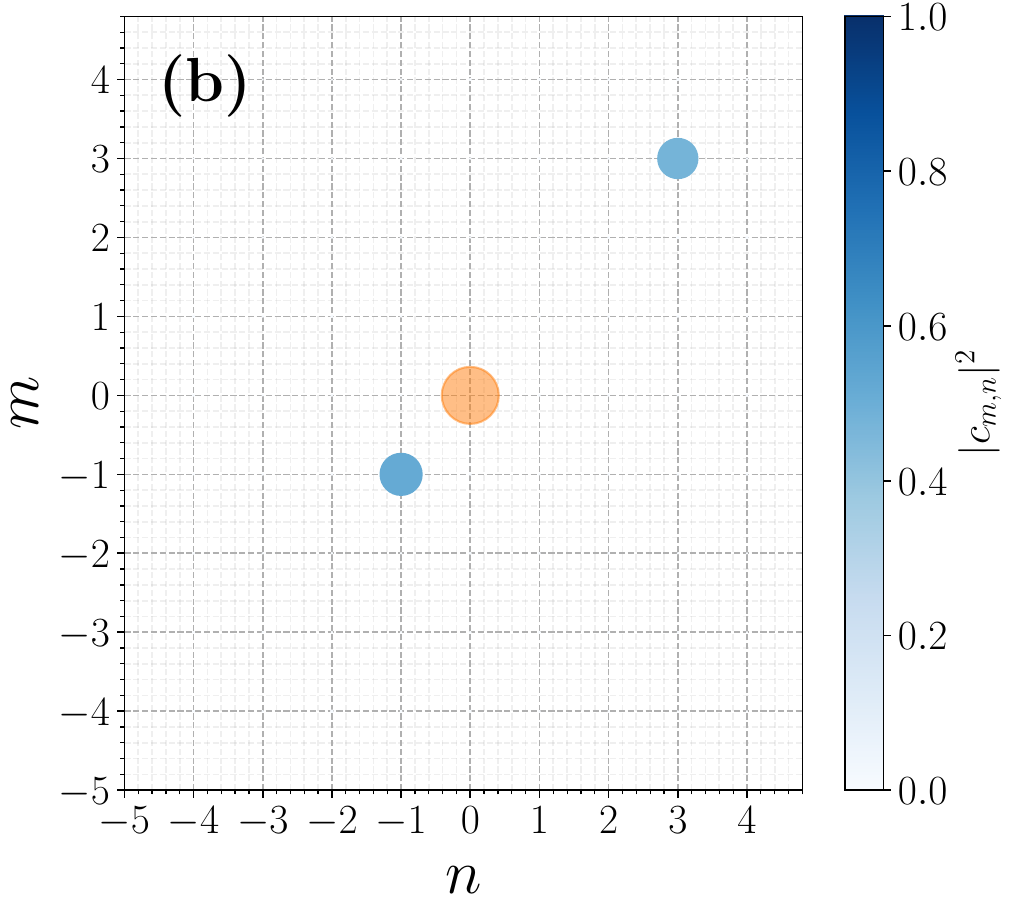}
        \caption{Same as Fig.~\ref{fig:2Dpopcont} but for the  target state $\frac{1}{\sqrt{2}}\left( |\phi_{-1,-1}\rangle + |\phi_{3,3}\rangle \right)$. Only the controls $\varphi_{3,1}$ and $\varphi_{2,3}$ are optimized.}
    \label{fig:2Dpopcont_ns}
\end{figure}

\begin{figure}
    \centering
     \includegraphics[width=0.5\textwidth]{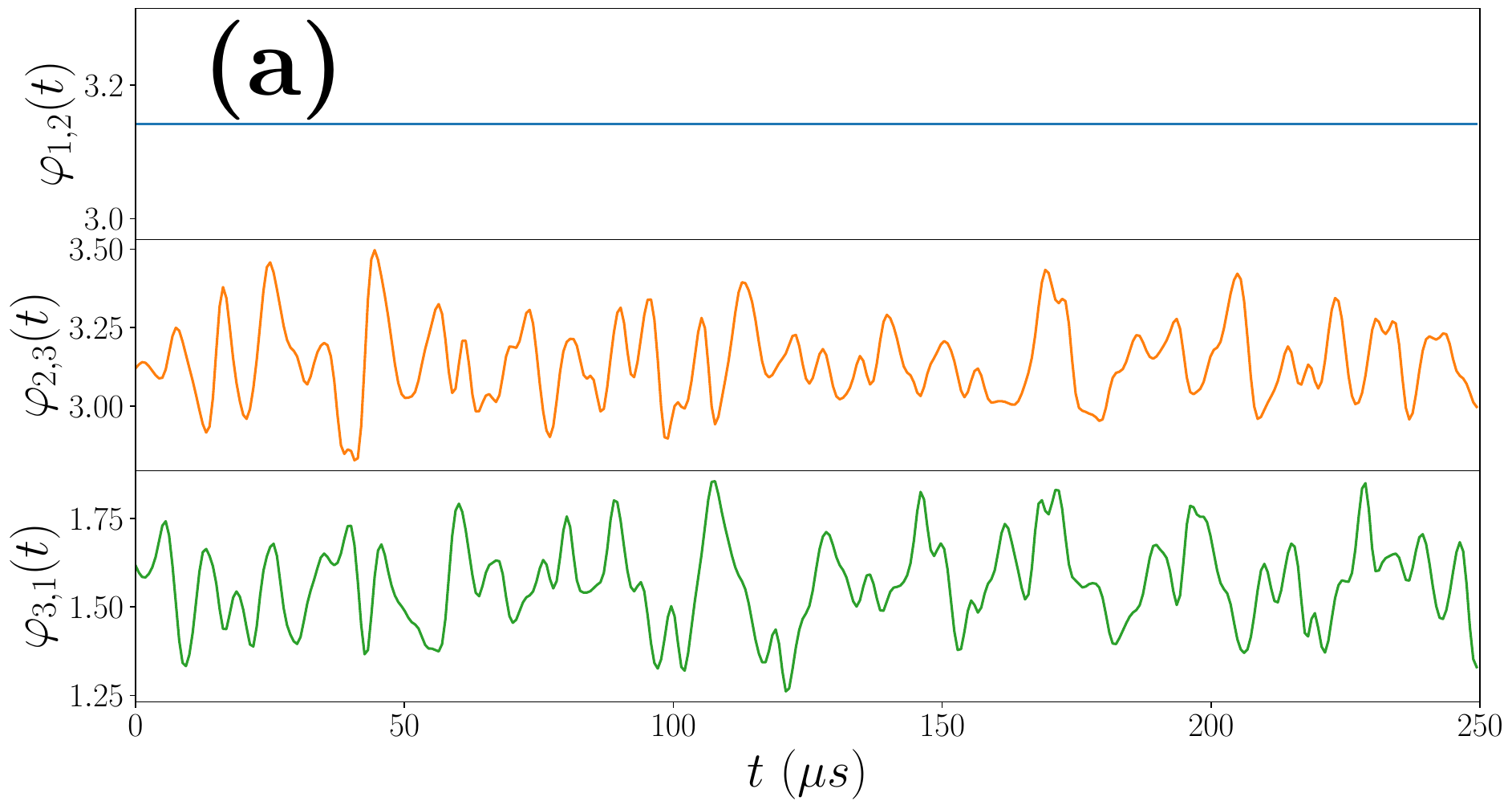}
     \includegraphics[width=0.5\textwidth]{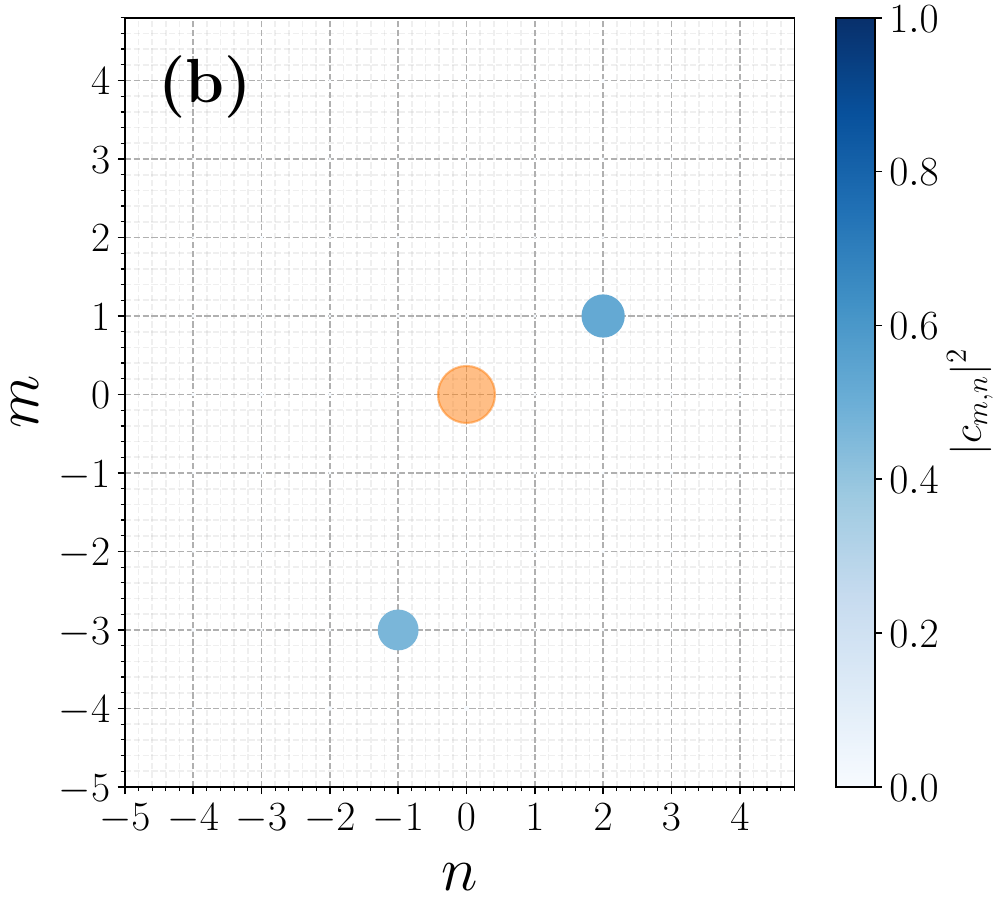}
        \caption{Same as Fig.~\ref{fig:2Dpopcont} but for the target state $\frac{1}{\sqrt{2}}\left( |\phi_{1,2}\rangle + |\phi_{-3,-1}\rangle \right)$. Only the controls $\varphi_{3,1}$ and $\varphi_{2,3}$ are optimized.}
    \label{fig:2Dpopcont_ns2}
\end{figure}

As an illustrative example, we consider the state-to-state transfer from the initial state $|\psi(0)\rangle =|0,0\rangle$ to a target state $|\psi_t\rangle=\frac{1}{\sqrt{2}}\left(|3,3\rangle + |-3,-3\rangle\right)$ with a control time $t_f = 250 \ \mu s$. The values of the other parameters are the same as in Sec.~\ref{sec2}.
Figure~\ref{fig:2Dpopcont} represents the optimization result for this control problem. Note that the controls are not independent since $\varphi_{12}$ and $\varphi_{31}$ are equal. This observation also illustrates the role of the number of controls for the controllability of this system. Figure~\ref{fig:2Dpopcont_ns} shows this point for another target state. In this case, only two controls $\varphi_{2,3}$ and $\varphi_{3,1}$ are optimized, the third one is kept constant in time. Despite this constraint, the target state is reached with a very good precision. Finally, in Fig.~\ref{fig:2Dpopcont_ns2}, we consider a non-symmetric state in $m$ and $n$ with $|\varphi_t\rangle=\frac{1}{\sqrt{2}}(|\phi_{1,2}\rangle+
|\phi_{-3,-1}\rangle)$. Again, our algorithm designs a very efficient control process with a fidelity above 0.99. We note that a broad range of states can be reached with just two variable phases, which raises the question of the controllability of this system, that is the relationship between the set of accessible states, the number of available controls and the lattice configuration.

\section{Conclusion and prospective views}\label{sec5}
We have proposed two extensions of the gradient-based algorithm, GRAPE, for controlling a BEC in an optical lattice. We have shown how to adapt this algorithm to the non-linear case where the atomic interaction is not neglected and for a two-dimensional optical lattice. The two generalizations are supported by a mathematical analysis of the optimal control problem based on the PMP. Numerical examples show the effectiveness of the different procedures in experimentally realistic cases. We emphasize that such algorithms have the advantage of simplicity and general applicability, regardless of the structure of the optical lattice at two or three dimensions or the number of available controls. Thus, based on the material presented in this paper, GRAPE can be adapted to other geometric configurations.

This work describes in detail how to numerically solve  the optimal control of a BEC in an optical lattice, but it also raises a number of issues that go beyond the scope of this paper. A first problem concerns the controllability of this system under  experimental constraints and limitations on pulse amplitude and duration. A general problem is to describe the set of reachable sets for a given lattice configuration and physical constraints on the control parameters. Some mathematical results have shown the small-time controllability of this system in an ideal situation~\cite{pozzoli2023,pozzoli2024}. However, such results do not use a directly implementable control strategy  and consider  pulses with a very large amplitude. Another interesting question is to evaluate the role of the nonlinearity on the optimal solution. It has been shown in~\cite{deffner2022} that the quantum speed limit~\cite{deffner2017} grows with the nonlinearity strength of the BEC. An intriguing question is to test this conjecture in our system using time-optimal control protocols. If this statement is verified, it could be of utmost importance to minimize the preparation time of specific states by playing with the nonlinearity of the BEC dynamics. This work assumes that the Hamiltonian parameters of the system are exactly known. It is not the case in practice where the experimental parameters are estimated within a given range. This limitation could be partly avoided by generalizing to this system the robust control pulses known for two-level quantum systems~\cite{kobzar2012,li2009,vandamme2017,lapert2012}. An optimal control can be designed in this case by considering the simultaneous control of a set of quantum systems characterized by a different value of the parameter. In this study, we only consider state-to-state transfer, but it will be interesting to extend this approach to quantum gates~\cite{palao2003,palao2002}. Recently, a rigorous foundation of quantum computing in the nonlinear case has been developed~\cite{xu2022}. In this direction, a first goal will be to show that the non-linear GRAPE algorithm described in this work can be used to implement quantum gates. Work is underway in our group to answer such questions both theoretically and experimentally.

\section*{Funding}
We thank the support from the Erasmus Mundus
Master QuanTeem (Project Number: 101050730), the project
QuanTEdu-France (ANR-22-CMAS-0001) on quantum technologies and the ANR project QuCoBEC (ANR-22-CE47-0008-02). This project has received financial support from the CNRS through the MITI interdisciplinary programs (QUSPIDE project).

\section*{References}
\bibliographystyle{vancouver}

\end{document}